\documentclass[floatfix,a4paper,aps,prd,twocolumn,preprintnumbers]{revtex4-1}  
\usepackage{epsfig} 
\usepackage{amsmath} 
\usepackage{amssymb} 
\usepackage{amsfonts} 
\usepackage{graphicx} 
\usepackage{graphics,subfigure} 
\usepackage{float} 
\usepackage{booktabs} 
\usepackage{color, ulem} 
\usepackage{geometry} 
\usepackage{rotating} 
\usepackage{units} 
\usepackage{slashed} 
 
\newcommand{\bea}{\begin{eqnarray}} 
\newcommand{\eea}{\end{eqnarray}} 
\newcommand{\beq}{\begin{equation}} 
\newcommand{\eeq}{\end{equation}}  
\newcommand{\beqa}{\begin{eqnarray}} 
\newcommand{\eeqa}{\end{eqnarray}} 
\newcommand{\bit}{\begin{itemize}} 
\newcommand{\eit}{\end{itemize}} 
\newcommand{\barr}{\begin{eqnarray}}
\newcommand{\earr}{\end{eqnarray}}

\newcommand{\met}{\ensuremath{\slashed{p}_T}} 
\newcommand{\mpt}{\ensuremath{\slashed{p}_T}} 
 
\newcommand{\spot}{\mathcal{W}}
\newcommand{\rpv}{{{\not\!{R}}_p}}
\newcommand{\rp}{R_p}
\newcommand{\bt}{\textrm{B}_3}
\newcommand{\eps}{\epsilon}

\newcommand{\lam}{\lambda}
\newcommand{\lamp}{\lambda'}

\newcommand{\sgnmu}{\textrm{sgn}(\mu)}

\newbox\charbox 
\newbox\slabox 
\def\s#1{{      
    \setbox\charbox=\hbox{$#1$} 
    \setbox\slabox=\hbox{$/$} 
    \dimen\charbox=\ht\slabox 
    \advance\dimen\charbox by -\dp\slabox 
    \advance\dimen\charbox by -\ht\charbox 
    \advance\dimen\charbox by \dp\charbox 
    \divide\dimen\charbox by 2 
    \raise-\dimen\charbox\hbox to \wd\charbox{\hss/\hss} 
    \llap{$#1$} 
}}

\begin{document} 
  
\title{Testing neutrino masses in the R--parity violating
 minimal supersymmetric standard model with LHC results} 
 
 
\author{M. Hanussek} 
\email[]{hanussek@th.physik.uni-bonn.de} 
\affiliation{Bethe Center for Theoretical Physics, University of Bonn, Bonn,  Germany} 
 
\author{J.~S. Kim} 
\email[]{jongsoo.kim@adelaide.edu.au} 
\affiliation{ARC Centre of
 Excellence for Particle Physics at the Terascale, School of
 Chemistry and Physics, University of Adelaide, Adelaide, Australia}

\begin{abstract}
  Within the R--parity violating minimal supersymmetric standard model
  (MSSM), we use a hierarchical ansatz for the lepton--number
  violating trilinear Yukawa couplings by relating them to the
  corresponding Higgs--Yukawa couplings. This ansatz reduces the
  number of free parameters in the lepton--number violating sector
  from 36 to 6.  Baryon--number violating terms are forbidden by
  imposing the discrete gauge symmetry Baryon Triality.  We fit the
  lepton--number violating parameters to the most recent neutrino
  oscillation data, including the mixing angle $\theta_{13}$ found by
  Daya Bay. We find that we obtain phenomenologically viable neutrino
  masses and mixings only in the case of normal ordered neutrino
  masses and that the lepton--number violating sector is unambiguously determined
  by neutrino oscillation data.  We discuss the resulting collider
  signals for the case of a neutralino as well as a scalar tau
  lightest supersymmetric particle. We use the ATLAS searches for
  multi--jet events and large transverse missing momentum in the 0, 1
  and 2 lepton channel with 7 TeV center--of--mass energy in order to
  derive exclusion limits on the parameter space of this R--parity
  violating supersymmetric model.
\end{abstract} 
\preprint{ADL-12-18-T785}
\maketitle 
  
\section{Introduction} 
\label{sec:intro} 
A main objective of both multi--purpose experiments ATLAS and CMS at
the Large Hadron Collider (LHC) is the search for new physics beyond
the Standard Model (SM). Many of these extensions, in particular
supersymmetry (SUSY) \cite{Haber:1984rc,Drees:2004jm}, include new
heavy colored states and a weakly interacting lightest new particle
escaping detection. Thus the most generic signal among these models
are several hard jets and large transverse missing momentum ($\met$).
ATLAS and CMS grouped their multi--jet and missing transverse momentum
searches into 0, 1, 2 lepton
studies~\cite{Aad:2011ib,Aad:2011qa,ConfNote0lept,CMS0lepta,CMS0leptb,
Chatrchyan:2011zy,ATLAS:2011ad,ConfNote1lept,CMS1lept,Aad:2011cwa,CMS2lepta,CMS2leptb},
in order to be sensitive to different SUSY models and to avoid an
overlap between these studies. Most studies were recently updated to
the full dataset of about 5 fb$^{-1}$ recorded in 2011 at a
center--of--mass energy of 7 TeV.  So far, no excess above SM
expectations has been observed and strict bounds on any supersymmetric
model or another relevant new physics model providing a similar collider
signal can be derived. ATLAS and CMS mainly concentrate on
SUSY searches which are based on R--parity conserving ($\rp$)
supersymmetric extensions of the SM \cite{Farrar:1978xj}. An equally
well motivated scenario is a R--parity violating ($\rpv$)
supersymmetric SM \cite{Dreiner:1997uz}, where the discrete symmetry
baryon triality ($\bt$) \cite{Dreiner:2005rd} is imposed in order to
avoid baryon--number violation and proton decay. The particle spectrum is the same as
for $\rp$ models. However, lepton (L--) number is violated and the lightest supersymmetric particle (LSP)
is not stable any more. Thus an alternative dark matter candidate may
be needed such as the axino or
gravitino~\cite{Kim:2001sh,Buchmuller:2007ui}.  In principle, any
supersymmetric particle can now be the lightest supersymmetric
particle (LSP) \cite{Dreiner:2009fi}.  The LSP decays lead to
observable effects at the LHC, which can be significantly different
from models with R--parity conservation
\cite{Desch:2010gi,Dreiner:2011wm}.  Also, the L--violation causes
massive neutrinos to emerge in the $\bt$ minimal supersymmetric SM
(MSSM)~\cite{Allanach:2007qc,Dreiner:2010ye,Dreiner:2011ft,Dreiner:2007uj}
without introducing a new see--saw mass scale or extending the
particle spectrum \cite{Mohapatra:1979ia,Minkowski:1977sc}. Data
from neutrino experiments can be used to constrain the L--violating
couplings \cite{Dreiner:2010ye}.

Within the $\bt$ minimal supersymmetric SM (MSSM), we make a
hierarchical ansatz in the L--violating sector, relating the trilinear
L--violating Yukawa couplings to the Higgs--Yukawa couplings, as first
proposed in Ref.~\cite{Dreiner:2007uj}.  This reduces the number of
free L--violating parameters to six.  We take into account
experimental results on neutrino oscillations, which amounts to five
constraints (neutrino mixing angles and mass-squared
differences). When additionally fixing the overall neutrino mass
scale, this enables us to unambiguously determine the magnitude of the
six L--violating parameters, removing all degrees of freedom from the
L--violating sector.

Consequently, the decay properties of the LSP in the hierarchical
$\bt$ MSSM depend only on the experimental neutrino data.  We expect
no difference in the production and decay chains of supersymmetric
particles compared to the $\rp$ MSSM, since the magnitude of the
L--violating couplings needs to be fairly small (of order $10^{-5}$)
in order to be in accordance with neutrino data.

There have been several ATLAS and CMS searches as well as
phenomenological studies for $\rpv$ models, based on resonant slepton
production, multi--lepton signatures or displaced
vertices~\cite{ConfNote4lept,
  ConfNote4leptRPV,Aad:2011qr,Aad:2011zb,Chatrchyan:2011ff,CMS:2012aa,graham:2012aa}.
However, most of these studies constrain models where the L--violating
couplings are either very large (for single slepton production), very
small (for displaced vertices) or where we have single coupling
dominance and four body decays (4 lepton
signature)~\cite{Dreiner:2012mn}. Neither of these criteria is the
case in most regions of the hierarchical $\bt$ MSSM parameter space.
Apart from these studies, the results of the ATLAS 1 lepton,
multi--jet and $\met$ study with $1$ fb$^{-1}$ of data were used to
restrict a bilinear R--parity violating model \cite{Hirsch:2000ef},
which takes into account constraints from neutrino data
\cite{ATLAS:2011ad}.
  
In this study, we would like to re--interpret the ATLAS studies with
jets, $\met$ and 0, 1 or 2 isolated leptons
\cite{Aad:2011ib,ATLAS:2011ad,Aad:2011cwa} in the light of the
hierarchical $\bt$ MSSM. Except for the 2 lepton study, which uses $1$
fb$^{-1}$, the studies have been updated to $5$ fb$^{-1}$
\cite{ConfNote0lept,ConfNote1lept}, using the full 2011 data. Since
in a generic $\bt$ MSSM, the number of free parameters in the SUSY
breaking sector is too large to perform a systematic study, we work in
the $\bt$ constrained MSSM ($\bt$ cMSSM) \cite{Allanach:2003eb}, which
imposes simplifying assumptions on the scalar and gaugino masses and
couplings at the unified (GUT) scale. It turns out that only specific
regions of the cMSSM parameter space are phenomenologically viable when
taking into account neutrino data~\cite{Dreiner:2010ye}, and we focus
on these parameter regions. As a result, there are 4 free
parameters in the SUSY breaking sector besides the six L--violating parameters.

In Sect.~\ref{b3_cssm} , we shortly discuss how neutrino masses are
generated in the hierarchical $\bt$ cMSSM. We then describe how we fit
the L--violating parameters in order to obtain the correct masses and
mixing angles of the neutrino sector at any parameter point in the
hierarchical $\bt$ cMSSM parameter space.  In
Sect.~\ref{sec:signal_description}, we examine the arising collider
signatures for the case of stau LSP and neutralino LSP scenarios.  In
Sect.~\ref{sec:lhc_analysis}, we present bounds on the hierarchical
$\bt$ cSSM neutrino model derived from SUSY ATLAS searches.  We
conclude in Sect.~\ref{sec:summary}.

\section{Hierarchical baryon triality cMSSM and massive neutrinos}
\label{b3_cssm}

\subsection{Hierarchical Baryon Triality ($\bt$) cMSSM}
 \label{sec:b3} 
 The $\bt$ MSSM allows for additional, L--violating terms in the superpotential
 compared to the $\rp$ MSSM~\cite{Weinberg:1979sa,Sakai:1981pk,Weinberg:1981wj},
\bea 
\spot_{\bt} &=& \spot_{\rp} + \eps_{ab}\, [\frac{1}{2}\lam_{ijk}
L_i^aL_j^b\bar{E}_k +
\lamp_{ijk} L_i^aQ_j^b\bar{D}_k \nonumber \\
&& \phantom{\eps_{ab}} - \kappa_i L_i^aH_u^b ].   \label{spotLNV}
\eea 
$L_i$, $Q_i$ correspond to the SU(2) doublet lepton and quark
superfields. $\bar E_i$, $\bar D_i$ are the SU(2) singlet lepton and
down--type quark superfields, respectively.  $i,j,k\,\in\{1,2,3\}$ are
generation indices, $a,b\in\{1, 2\}$ ($\eps_{12}=1$) are indices of
the $SU(2)_L$ fundamental representation, while the corresponding
$SU(3)_c$ indices are suppressed. The trilinear couplings
$\lambda_{ijk}$ correspond to nine independent parameters due to the
antisymmetry of the first two indices $i,j$, whereas the trilinear
couplings $\lambda_{ijk}^\prime$ denote 27 independent parameters.
The bilinear couplings $\kappa_i$ are 3 dimensionful couplings.  

For universal supersymmetry breaking, the bilinear L--violating
couplings and the corresponding soft--breaking terms can be
simultaneously rotated to zero at the unification (GUT) scale via a
basis transformation of the lepton and Higgs superfields
\cite{Allanach:2003eb,Dreiner:2003hw}.  However, non--vanishing
$\kappa_i$ terms (and non--aligned soft--breaking terms) are generated
at the electroweak scale via the renormalization group
equations~\cite{Nardi:1996iy}.

In the $\bt$ constrained MSSM (cMSSM), the number of free parameters
in the soft--breaking sector is constrained. We end up with $5+n$
independent parameters at the GUT scale \cite{Allanach:2003eb},
\beq
 M_0,\,M_{1/2},\,A_0,\, \sgnmu,\,\tan{\beta},\,\Lambda. 
\label{B3CMSSM}
\eeq
$M_0$, $M_{1/2}$ and $A_0$ denote the universal scalar mass, universal
gaugino mass and universal trilinear scalar coupling,
respectively. $\sgnmu$ is the sign of the superpotential Higgs mixing
parameter and $\tan\beta$ is the ratio between the two Higgs vacuum
expectation values. $\Lambda$ denotes a subset of $n$ independent
dimensionless trilinear L--violating couplings.

In this work, we further restrict the number of free L--violating
parameters: In the $\bt$ cMSSM, the down--type Higgs superfield and
the SU(2) doublet lepton superfield have the same gauge quantum
numbers \cite{Hall:1983id}. They are indistinguishable because lepton
number is broken. Thus, the L--violating trilinear terms in
Eq.~(\ref{spotLNV}) resemble terms in the R--parity conserving
superpotential,
\bea
\spot_{\rp} &\supset& \eps_{ab} \,[(Y_E)_{jk}H_d^aL_j^b\bar{E}_k +
(Y_D)_{jk}H_d^aQ_j^b\bar{D}_k], 
\eea
where $(Y_E)_{jk}$ and $(Y_D)_{jk}$ are the Higgs--Yukawa couplings of
the lepton and the down--type quarks, respectively. We therefore
proposed the following ansatz at the GUT scale~\cite{Dreiner:2007uj},
which can be motivated in the framework of Froggatt-Nielsen models \cite{Dreiner:2003hw}
\begin{eqnarray}
\lam_{ijk}&\equiv&\ell_{i}\cdot \left(Y_E\right)_{jk}~-~\ell_{j}\cdot \left(Y_E\right)_{ik}
\label{ansatz_lam}\,,
\\
\lam^\prime_{ijk}&\equiv&\ell^\prime_{i}\cdot \left(Y_D\right)_{jk}
\label{ansatz_lamp}\,.
\end{eqnarray}
Here, $\ell_i,\,\ell^\prime_i$ are $c$-numbers. Eq.~(\ref{ansatz_lam})
has the required form to maintain the anti-symmetry of the
$\lam_{ijk}$ in the first two indices. Assuming a specific form of the
Higgs--Yukawa couplings, the number of L--violating parameters reduces
to six complex numbers. We have given our ansatz in the weak--current
basis. However, after EW symmetry breaking, we must rotate to the
mass--eigenstate basis. Experimentally, only the PMNS and the CKM
matrix are known \cite{Cabibbo:1963yz,Kobayashi:1973fv}. The explicit
lepton and quark mixing matrices are therefore not fully
determined. In the following, we assume that the lepton Higgs-Yukawa
matrix is diagonal
Thus, we assume mixing only in the neutrino
sector for the leptonic sector. In the quark sector, we assume
left--right symmetric mixing. Additionally, we work in the limit where
the down--type Higgs--Yukawa matrix is diagonal whereas the up--type
is non--diagonal. Hence our specific form of the Higgs--Yukawa
couplings implies mixing only in the up--type--sector. In
  Ref.~\cite{Dreiner:2011ft}, it was shown that the choice of quark mixing
  (e.g. mixing in the up--type versus mixing in the
  down--type--sector) does not significantly influence the numerical
  results at the low energy scale.
\subsection{$\bt$ neutrino masses}\label{B3neutrinoMasses}
Since lepton number is violated, the neutrinos mix with the
neutralinos, resulting in a 7x7 neutralino-neutrino mass matrix of
rank 5.  As a result, we obtain one massive neutrino at tree
level~\cite{Allanach:2003eb},
\beq
m_\nu^{\rm{tree}}=- \frac{16 \pi \alpha_{\textrm{GUT}}}{5} \: 
\frac{\sum_{i=1}^{3}  \left(v_i-v_d\frac{\kappa_i}{\mu}\right)^2}
{M_{1/2}}
\eeq
Here $v_d$, $v_u$ and $v_i$ denote the vacuum expectation values of
the $H_d$, $H_u$ and sneutrino fields.  However, experimental neutrino
oscillation data suggests that we need at least two massive neutrinos.
Since there is only one massive neutrino at tree--level, higher--order
corrections need to be taken into account.  Full 1--loop corrections
to the neutrino-neutralino mass matrix have been discussed in
Ref.~\cite{Dreiner:2011ft}. A good estimate of the size of these
radiative corrections is given by the slepton--lepton and down--type
quark--squark loop contribution, which are proportional to
\cite{Grossman:1998py}
\bea
\left(m_\nu^{\ell}\right)_{ij}& \propto&\lambda_{ikn}\lambda_{jnk} 
\: m_{\ell_k}m_{\ell_n} ,\\
\left(m_\nu^{d}\right)_{ij}& \propto &N_c \;
\lambda^\prime_{ikn}\lambda^\prime_{jnk} \: 
m_{d_k}m_{d_n}.
\eea 
The proportionality of the loop contributions to the exchanged SM fermion
mass in the loop further increases the effect that trilinear couplings
with indices $i33$ are dominant over all other indices $ijk$, as is
clear from the hierarchical ansatz in Eqs.~(\ref{ansatz_lam}) and
(\ref{ansatz_lamp}).

Ref.~\cite{Dreiner:2010ye} noted that in large regions of cMSSM
parameter space the ratio between the tree--level neutrino mass and
the radiative contributions is too large too yield a
phenomenologically viable neutrino mass hierarchy. However, due to RGE
effects in the running of L-violating parameters, the tree--level
neutrino mass has a global minimum at
\bea
A_0^{(\lambda^\prime)}&\approx&2 M_{1/2},
\label{A0_minimum_lp}\\
A_0^{(\lambda)}&\approx&\frac{M_{1/2}}{2},
\label{A0_minimum_l}
\eea
for non--zero $\lambda^\prime_{ijk}$ or $\lambda_{ijk}$, respectively.
We choose $A_0$ such that it minimizes the $\lambda^\prime$
contribution to neutrino masses [Eq.~(\ref{A0_minimum_lp})], as
explained in more detail in the next
paragraph.
Thus, in the hierarchical $\bt$ cMSSM a set of 10 free parameters,
\beq
M_{1/2},\,M_{0},\,\sgnmu,\,\tan\beta,\,\ell_i,\,\ell_i^\prime,
\label{b3_cmssm}
\eeq 
fixes the full $\bt$ cMSSM.  

As described in Ref.~\cite{Dreiner:2011ft}, it is possible to 
obtain the experimentally measured neutrino mass squared differences and
mixing angles by independently generating
each neutrino mass with a set of three L--violating free parameters.
This means that 6 or 9 independent couplings are
necessary in order to obtain the full spectrum with either two or
three massive neutrinos. However, in the case of neutrinos in normal
hierarchy mass ordering with a massless lightest neutrino, it turns
out that one can do with only 2 couplings to explain the heaviest
neutrino mass, $m_{\nu_3}$, cf. Ref.~\cite{Dreiner:2011ft}.  This is fortunate, 
because due to our
hierarchical ansatz only $\ell_i^\prime$, $\ell_1$ and $\ell_2$ have a
significant impact on the neutrino sector whereas $\ell_3$ generates
only a negligible contribution to the neutrino masses if it is of the same
order of magnitude as the other couplings~\cite{foot1}. Therefore, we generate
$m_{\nu_3}$ at tree--level via the $\lam_{ijk}$ couplings, which are
in turn determined by $\ell_1$ and $\ell_2$.  The second neutrino
mass, $m_{\nu_2}$ is generated via $\lamp_{ijk}$ (determined by the
$\ell'_i$) at one--loop level, whereas the lightest neutrino must
remain massless, $m_{\nu_1} \approx 0$.
 
In summary, we have 5 free L-violating parameters which control the
neutrino sector, $\ell_i^\prime$ and $\ell_1$, $\ell_2$. These can be
used to generate non-zero $m_{\nu_2}$ and $m_{\nu_3}$, respectively,
in accordance with the two mass squared difference and three mixing
angles from experiment. It is not easily possible to obtain inverse
hierarchy or degenerate neutrino masses in the hierarchical $\bt$
cMSSM unless $\ell_3$ becomes several orders of magnitude larger than
the other L-violating parameters. 

\subsection{Experimental neutrino oscillation data}
Assuming three active oscillating neutrinos, the best global fit
values of the neutrino masses and mixing parameters at
$1\sigma$ C.L. are given by~\cite{Schwetz:2011qt,An:2012eh},
\bea
\sin^2[\theta_{12}]&=&0.31\pm0.02,\nonumber \\
\sin^2[\theta_{23}]&=&0.51\pm0.06,\nonumber \\
\sin^2[2\theta_{13}]&=&0.09\pm0.02,\nonumber \\
\Delta m_{21}^2&=&7.59\pm0.2\times10^{-5}\,\rm{eV^2},\nonumber \\
\Delta m_{31}^2&=& \left\{
\begin{array}{r}
-2.34\pm0.1\times 10^{-3}\textrm{ eV}^2\\
2.45\pm0.1\times 10^{-3}\textrm{ eV}^2
\end{array} \right\}
\label{best_fit}
\eea
where 
\beq
\Delta m_{ij}^2\equiv m_{\nu_i}^2-m_{\nu_j}^2.
\eeq
$m_{\nu_i}$ denote the neutrino masses in order of largest
electron-neutrino admixture.  There are two large mixing angles
$\theta_{12}$ and $\theta_{23}$. Deviating from
  Ref~\cite{Schwetz:2011qt}, we use in Eq.~(\ref{best_fit}) for
  $\theta_{13}$ the best fit value recently measured by Daya Bay and
  RENO ~\cite{An:2012eh, Ahn:2012nd}. The neutrino oscillation data
implies at least two non--vanishing neutrino masses $m_{\nu_i}$. In
this work, we usually consider the so--called Normal Hierarchy (NH)
scenario, where $\Delta m_{31}^2>0$ and $m_{\nu_1}\!\approx0$.

\subsection{Numerical results}\label{NumResults}
For each cMSSM point we fit the L--violating parameters $\ell_i$ and
$\ell_i^\prime$ to the best--fit Normal Hierarchy neutrino mass data in
Eq.~(\ref{best_fit}).  We perform this fit by minimizing the $\chi^2$
function
\beq
\chi^2=\frac{1}{N_{\rm obs}}\sum_{i=1}^{N_{\rm obs}}
\left(\frac{f_i^{\rm softsusy}-f_i^{\rm obs}}{\delta_i}\right),
\eeq
where $f_i^{\rm obs}$ are the central values of the $N_{\rm obs}$ experimental
observables in Eq.~(\ref{best_fit}), $f_i^{\rm softsusy}$ are the
corresponding numerical predictions and $\delta_i$ are the $1\sigma$
uncertainties. We calculate the low energy mass spectrum and couplings
with {\tt SOFTSUSY3.3.0} \cite{Allanach:2011de}. The numerical
minimization of our $\chi^2$ function is done with the program
package {\tt MINUIT2} \cite{James:1975dr}. Details of our numerical
procedure can be found in Ref.~\cite{Dreiner:2011ft}. Here, we present
an example solution where we translate the best fit values $\ell_i$
and $\ell_i^\prime$ into the corresponding values of the trilinear
L-violating couplings at the unification scale:
\barr
\lambda_{133}&=&1.72\cdot10^{-6} \nonumber \\
\lambda_{233}&=&2.74\cdot10^{-6}\nonumber \\
\lambda_{133}^\prime&=&1.13\cdot10^{-5}\nonumber \\
\lambda_{233}^\prime&=&3.89\cdot10^{-5}\nonumber \\
\lambda_{333}^\prime&=&3.11\cdot10^{-5} \label{solBP1}
\earr

\begin{center}
\begin{figure}
\includegraphics[scale=0.6]{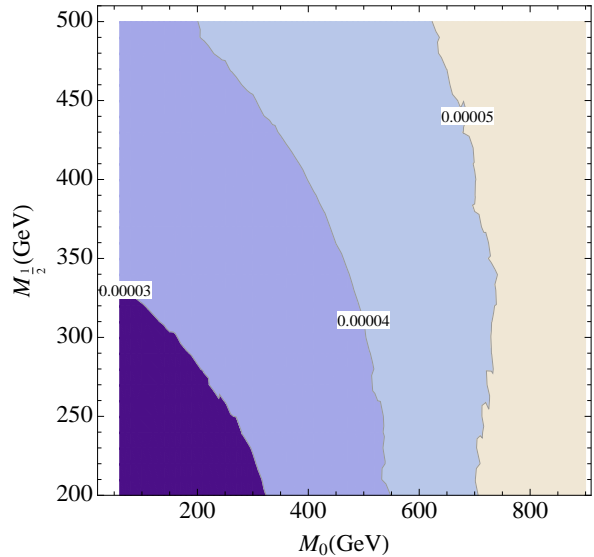}
\caption{Best--fit values of the L--violating coupling
  $\lamp_{233}$ at the unification scale in the
  $M_0$--$M_{1/2}$ plane, fixing
  $A_0^{(\lamp)} \approx 2 M_{1/2}$, $\tan\beta=25$ and $\sgnmu=+1$.}\label{lampi33}
\end{figure}
\end{center}

We have used $M_0=100$ GeV, $M_{1/2}=500$ GeV, $\tan\beta=25$, $\sgnmu$ and
$A_0^{(\lamp)} \approx 2 M_{1/2}$.  As one can see, the
$\lambda_{i33}$ and $\lambda_{i33}^{\prime}$ couplings are between
$\mathcal{O}(10^{-5})$ and $\mathcal{O}(10^{-6})$. All remaining
trilinear L--violating couplings are at least one order of magnitude
smaller, below $\mathcal{O}(10^{-7})$. The couplings
$\lambda_{233}^\prime$ and $\lambda_{333}^\prime$ tend to be the
largest trilinear L--violating couplings.  In Fig.~\ref{lampi33}, we
display the best fit value of $\lambda_{233}^\prime$ in the
$M_0$--$M_{1/2}$ plane. We see that the magnitude of the L--violating
couplings does not strongly depend on $M_0$ and
$M_{1/2}$. Furthermore, the relative magnitude of the L--violating
couplings to each other remains roughly the same throughout the
parameter space.

Recall that the parameter $\ell_3$ is not fixed by the neutrino
oscillation data in the normal hierarchy scenario.  However, we assume
that $\ell_3$ is of the same order of magnitude as $\ell_1$ and
$\ell_2$, setting $\ell_3=\ell_2$ in the rest of our
paper~\cite{foot2}.

We have checked all low energy constraints on the L--violating
trilinear couplings~\cite{Barbier:2004ez,Dreiner:2006gu}. However, in our case the
couplings are too small to have an observable impact on any low energy
observables.

\section{Collider signatures} 
\label{sec:signal_description} 
In this section, we investigate possible collider signatures of the
hierarchical $\bt$ cMSSM at the LHC. The best--fit values of the
L--violating couplings to neutrino data are too small to have an
observable effect on the resonant production of supersymmetric
particles. Thus, pair production of colored sparticles via strong
interactions is the dominant production channel at the LHC.  Only if
sleptons and gauginos are much lighter than the colored sparticles,
their production rate becomes comparable.
The produced sparticles cascade decay into the LSP.
 In our parameter space, we can have either a
stau LSP or a neutralino LSP~\cite{foot3}.  The final state collider signature is
determined by the decay properties of the LSP candidate.
 In the $\bt$
cMSSM, the LSP is almost always short--lived and decays within the detector via the
L-violating interactions~\cite{foot4}.
 We now describe the final
state signatures of stau LSP and neutralino LSP scenarios separately
after describing the numerical tools used. Then we go on to discuss in
which regions of $M_0-M_{1/2}$ parameter space they occur.

\subsection{Numerical tools}
\label{subsec:tools}
The low energy mass spectrum and couplings are calculated with {\tt
  SOFTSUSY3.3} \cite{Allanach:2011de}. The decay widths of the
relevant sparticles are obtained with {\tt
  IsaJet7.64}~\cite{Paige:2003mg} and {\tt IsaWig1.200}. However, the
decay channels of the neutralino LSP via the sneutrino vevs and the
$\kappa_i$ term are absent in {\tt IsaWig1.200}. Therefore, we
calculate decays via the bilinear L--violating couplings with {\tt
  SPheno3.1}~\cite{Porod:2011nf}. We combined all decay widths in
order to calculate the branching ratio of the sparticles. We use the
parton distribution functions MRST2007 LO
modified~\cite{Sherstnev:2007nd}. Our signal events are generated with
{\tt Herwig6.510}~\cite{Corcella:2000bw}. The cross sections are
normalized with the NLO calculations from {\tt Prospino2.1}~
\cite{Beenakker:1996ed} assuming equal renormalization and
factorization scale. Our events are stored in the Monte Carlo event
record format {\tt StdHep5.6.1}. We take into account detector effects
by using the fast detector simulation {\tt Delphes1.9} \cite{Ovyn:2009tx}, where we
choose the default ATLAS--like detector settings. Our event samples
are then analyzed with the program package {\tt ROOT}
\cite{Brun:1997pa} and we calculate the $95\%$ and $68\%$ confidence
levels (CL) of the exclusion limits with TRolke \cite{Rolke:2004mj}.

\subsection{Stau LSP decay}\label{stauSignatures}
In the parameter region where the lighter stau $\tilde \tau_1$ is the LSP, pair produced
squarks and gluinos at the LHC cascade decay into the LSP, 
producing jets and taus (tau--neutrinos) along the way,
\beq
pp \rightarrow \tilde q \tilde q/\tilde q \tilde g/\tilde g \tilde g  \rightarrow  \tilde \tau_1\tilde \tau_1+ 2j + X,
\label{prod_stau_lsp}
\eeq
where $j$ and $X$ denote jets and additional particles of the
process (such as $\tau$ or $\nu_{\tau}$), respectively. 
Note that we can have more than 2 jets in the final state if the process involves gluinos.
 These additional jets are included in $X$, which we discuss in more detail in Sect~\ref{subsect:scan}.
For example, right-handed squarks decay into a jet and the lightest
neutralino, which then typically decays into a stau and a tau with a
branching ratio of one,
\beq
\tilde q_R\, \tilde q_R\rightarrow j j \tilde
\chi_1^0\tilde \chi_1^0 \rightarrow j j\; \tau\tau \; \tilde \tau_1\tilde
\tau_1\,.
\label{prod_stau_lsp2}
\eeq

\begin{center}
\begin{figure}
\includegraphics[scale=0.8]{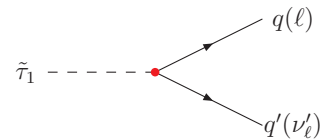}
\caption{Schematic characterization of the stau LSP decay in the hierarchical $\bt$ cMSSM.}
\label{FeynmanStau}
\end{figure}
\end{center}

The stau then directly decays into two SM fermions via the trilinear
L--violating couplings
$\lambda_{133}$, $\lambda_{233}$ and $\lambda_{3jk}^\prime$, cf. Fig~\ref{FeynmanStau}.
Decays via the $\lam_{i33}$ couplings are dominant, even though the
decay width via $\lamp_{3jk}$ is enhanced by a factor of $N_C=3$ and
the $\lambda_{3jk}^\prime$ couplings are generally larger. However,
the lightest stau is mostly right--handed and thus the coupling of the
stau via $\lamp$ is suppressed due to the small admixture with the
left--handed stau. Additionally, the stau decay via
$\lambda_{333}^\prime$ into a top and bottom quark is kinematically
forbidden or suppressed in large regions of parameter space. Stau
decays via $\lambda_{311}^\prime$ and $\lambda_{322}^\prime$ are
heavily suppressed due to the smallness of the couplings.

In principle, the stau can also mix with the charged Higgs boson via
$\kappa_3$ and decay via the two--body decay mode
$\tilde\tau\rightarrow\tau\nu$. However, we have numerically checked
that stau decays via bilinear operators are always sub--dominant in our model. 
We define a 
\bit
\item \textbf{benchmark point BP1} in the stau LSP region with
\vspace{-0.15cm} 
\begin{center}
$M_0=100$ GeV, $M_{1/2}=500$ GeV,  $\tan\beta=25$, $\sgnmu=1$ and $A_0^{(\lamp)} \approx 2 M_{1/2}$
\end{center}
\eit
This benchmark point is characterized by lightest neutralino, lighter stau, gluino and squark masses of 205 GeV, 162 GeV, 1146 GeV and 1012 GeV, respectively. 
 The dominant LSP branching ratios for {\bf BP1} are given by
\barr
\rm{Br}(\tilde\tau_1^-\rightarrow \tau^-\nu_e)&=&0.26\nonumber \\
\rm{Br}(\tilde\tau_1^-\rightarrow \tau^-\nu_\mu)&=&0.21\nonumber \\
\rm{Br}(\tilde\tau_1^-\rightarrow e^-  \nu_\tau)&=&0.26\nonumber \\
\rm{Br}(\tilde\tau_1^-\rightarrow \mu^- \nu_\tau)&=&0.21\nonumber \\
\rm{Br}(\tilde\tau_1^-\rightarrow s \: \bar c\:)&=&0.04 \;.
\earr
Note that the branching ratios into different decay channels are
roughly independent of the stau mass as long as the final state masses
are negligible. 

Roughly half of the staus decay into a charged lepton and neutrino, the
other half decays into a tau and neutrino.  Note that we only denote
electrons or muons as leptons in this paper.  Since one third of taus
decays leptonically, 
we expect final state collider signatures with either 0, 1 or 2
leptons from the decaying stau LSPs, for $12\%$, $46\%$ and $42\%$ of
events, respectively:
\barr
0\ell+2\nu+2\tau_{\rm had}+ 2j + X\nonumber \\
1\ell+2  (4) \nu+1\tau_{\rm had}+2j +X\nonumber \\
2\ell+2 (4,6) \nu+2j +X
\label{stau_final_signature}
\earr
where $\ell$ denotes an electron or muon and $\tau_{\rm had}$ denotes a
hadronically decaying tau.  If the lepton[s] in the $1\ell$ or $2 \ell$ channel come from 
a leptonically decaying tau, the number of neutrinos increases from 2 to 4 [6],
as shown in brackets in Eq.~(\ref{stau_final_signature}).  Due to the Majorana character of the
neutralino, both neutralinos can decay into like--charged staus
and hence we can have same--sign leptons in the final state.

\subsection{Neutralino LSP decay}
\begin{center}
\begin{figure}
\includegraphics[scale=0.8]{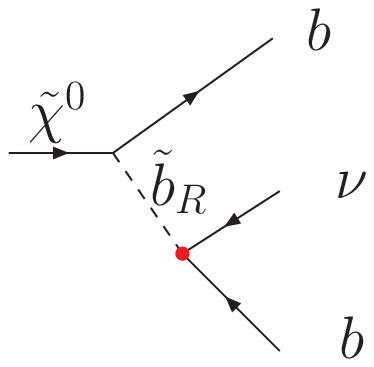} \hspace{0.74cm}
\includegraphics[scale=0.8]{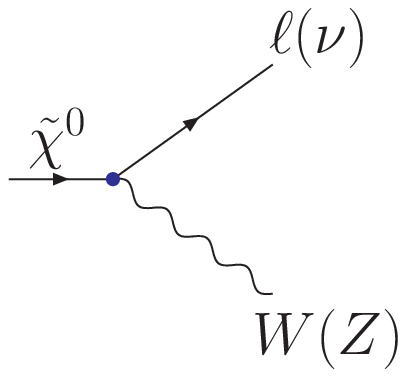}
\caption{Schematic characterization of the three--body (left) and two--body (right) decay modes of the neutralino LSP in the hierarchical $\bt$ cMSSM.}
\label{FeynmanChi}
\end{figure}
\end{center}

In the hierarchical $\bt$ cMSSM, the lightest neutralino eigenstate is
generally bino--like. The production process is given by
\beq
p p \rightarrow \tilde q \tilde q/\tilde q \tilde g/\tilde g \tilde g  \rightarrow \tilde
\chi_1^0\tilde \chi_1^0 + 2 j + X\,.\label{prod_chi_lsp}
\eeq
The neutralino LSP can either decay via a trilinear L--violating
operator into three SM fermions or via neutralino--neutrino mixing
(proportional to the bilinear L--violating couplings and the sneutrino
vevs) into a gauge/Higgs boson and a lepton,
cf. Fig.~\ref{FeynmanChi}.

For relatively low sfermion masses in the propagator, the trilinear
three--body decay modes dominate because the bilinear L--violating
couplings are only generated radiatively via RGE running and the
sneutrino vevs are determined to be relatively small from radiative
electroweak symmetry breaking.  However, in parameter regions with
heavy sfermions, the bilinear two--body decay mode becomes dominant
because the three body decay mode suffers from phase space suppression
and heavy virtual sfermions in the propagator.

First, we discuss the case where the lightest neutralino dominantly
decays via the trilinear LNV couplings, for which we define
\bit
\item \textbf{benchmark point BP2} with
\vspace{-0.15cm}
\begin{center}
$M_0=200$ GeV, $M_{1/2}=400$ GeV,  $\tan\beta=25$, $\sgnmu=1$ and $A_0^{(\lamp)} \approx 2 M_{1/2}$
\end{center}
\eit
This benchmark point is characterized by lightest neutralino, lighter stau, gaugino and squark masses of 163 GeV, 213 GeV, 937 GeV and 846 GeV, respectively. 
We obtain the following LSP branching ratios for {\bf BP2}:
\barr
\rm{Br}(\tilde\chi_1^0\rightarrow \overset{\textrm{\tiny(\textbf{--}\tiny)}}{\nu}_{\!\!\ell} b\bar b)&=&0.31\nonumber \\
\rm{Br}(\tilde\chi_1^0\rightarrow \overset{\textrm{\tiny(\textbf{--}\tiny)}}{\nu}_{\!\!\tau} b\bar b)&=&0.20\nonumber \\
\rm{Br}(\tilde\chi_1^0\rightarrow  W^\pm \ell^\mp)&=&0.21\nonumber \\
\rm{Br}(\tilde\chi_1^0\rightarrow  W^\pm \tau^\mp)&=&0.05\nonumber \\
\rm{Br}(\tilde\chi_1^0\rightarrow \overset{\textrm{\tiny(\textbf{--}\tiny)}}{\nu}_{\!\!\tau} Z^0)&=&0.13\nonumber \\
\rm{Br}(\tilde\chi_1^0\rightarrow \overset{\textrm{\tiny(\textbf{--}\tiny)}}{\nu}_{\!\!\tau} h^0)&=&0.08
\earr
The branching ratio of the three--body decay modes (the
$\tilde\chi_1^0\rightarrow \nu b \bar b$ channel) is roughly
51$\%$. However, for this benchmark point the two--body L--violating
decays via bilinear L--violating couplings already have a sizable
contribution to the LSP decays. 
The electron (electron-neutrino) channel is suppressed compared to the
muon decay channel because $\lamp_{133}\sim 0.3 \lamp_{233}$,
cf. Eq.~(\ref{solBP1}). Therefore, about $90\%$ of our leptons are
muons.  Summing up the various decay channels and including the gauge boson 
branching ratios, roughly $72\%$ of
neutralinos decay without leptons, $19\%$ with one lepton and $7\%$
with two leptons. This leads to $52\%$, $27\%$ and $14\%$ of events
with 0, 1 and 2 leptons from LSP decays, respectively. 

Assuming the cascade decay processes of Eq.~(\ref{prod_chi_lsp}),
dominant final state signatures are then given by
\barr
0\ell+2\nu+2 b \bar b+2j+X\nonumber \\ 
1\ell +1\nu + \,b\bar b+W_{\rm had}+2j+X\nonumber \\ 
2\ell +2\nu + \,b\bar b+2j+X \label{decaysBP2}
\earr

Next, we discuss the decay properties of the lightest neutralino in a region where the two--body decays dominate,
\bit
\item \textbf{benchmark point BP3} with
\vspace{-0.15cm}
\begin{center}
$M_0=600$ GeV, $M_{1/2}=400$ GeV,  $\tan\beta=25$, $\sgnmu=1$ and $A_0^{(\lamp)} \approx 2 M_{1/2}$
\end{center}
\eit
The lightest neutralino, lighter stau, gluino and squark masses of {\bf BP2} are 164 GeV, 579 GeV, 961 GeV and 1010 GeV, respectively. 
Here, the LSP decay channels are the same as for {\bf BP2}; however, the branching ratios differ drastically:
\barr
\rm{Br}(\tilde\chi_1^0\rightarrow \overset{\textrm{\tiny(\textbf{--}\tiny)}}{\nu}_{\!\!\ell} b\bar b)&=&0.04\nonumber \\
\rm{Br}(\tilde\chi_1^0\rightarrow \overset{\textrm{\tiny(\textbf{--}\tiny)}}{\nu}_{\!\!\tau} b\bar b)&=&0.03\nonumber \\
\rm{Br}(\tilde\chi_1^0\rightarrow  W^\pm\ell^\mp)&=&0.40\nonumber \\
\rm{Br}(\tilde\chi_1^0\rightarrow  W^\pm\tau^\mp)&=&0.14\nonumber \\
\rm{Br}(\tilde\chi_1^0\rightarrow \overset{\textrm{\tiny(\textbf{--}\tiny)}}{\nu}_{\!\!\tau} Z^0)&=&0.27\nonumber \\
\rm{Br}(\tilde\chi_1^0\rightarrow \overset{\textrm{\tiny(\textbf{--}\tiny)}}{\nu}_{\!\!\tau} h^0)&=&0.12
\earr
Since here the scalar masses ($M_0$) are fairly large, the two--body
neutralino decay modes via bilinear L--violating couplings or
sneutrino vevs dominate, amounting to $93\%$.  Therefore, there are
only half as many neutralinos decaying into the $0\ell$ channel as for
{\bf BP2}; twice as many decay into the $1\ell$ and $2\ell$ channel.
This results in final state signatures with 0,1 or 2 leptons at 24, 37
and 27$\%$, respectively.  Typical final state signatures
are given by
\barr
0\ell+2\nu+2 Z^0_{\rm had/\nu\nu}+2j+X\nonumber \\ 
1\ell +1\nu + \,Z^0_{\rm had/ \nu\nu}+W_{had}+2j+X\nonumber \\ 
2\ell +2\nu + \,Z^0_{\rm had/ \nu\nu}+2j+X \label{decaysBP3}
\earr
As mentioned before, the electron decay channel is
suppressed by roughly a factor of 10 compared to the muon decay channel.
 Additionally to the channels mentioned in Eq.~(\ref{decaysBP3}), there are $12\%$ of events with 3 or 4 leptons from LSP decay.

\subsection{Scan in the $M_0-M_{1/2}$ plane and kinematical distributions}\label{subsect:scan}

\begin{center}
\begin{figure}
\includegraphics[scale=0.35]{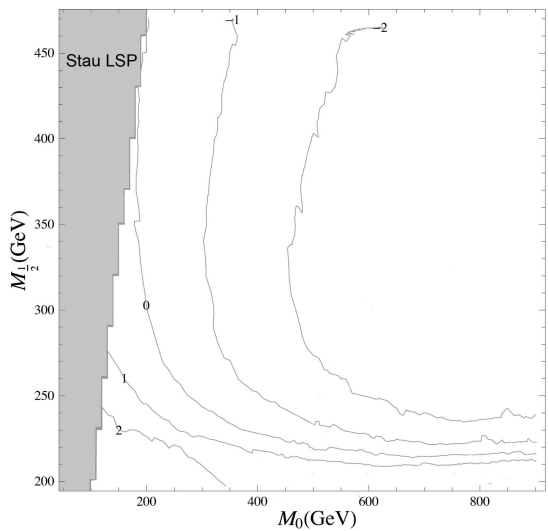}
\caption{The iso curves show the logarithmic ratio between three--body and two--body decay
  modes of the neutralino LSP in our benchmark region. In the stau LSP
  region, the two--body stau decay modes via the trilinear RPV couplings are
  always dominant.}
\label{TriBi}
\end{figure}
\end{center}

In the subsequent numerical analysis, we perform a scan in the
$M_0-M_{1/2}$ plane. For this, we define a benchmark region (BR) which
contains the three benchmark points defined above ({\bf BP1}, {\bf BP2}, {\bf BP3}):
\bit
\item \textbf{Benchmark region BR} (where $M_0$, $M_{1/2}$ free):
\begin{center}
  $\tan\beta=25$, $\sgnmu=1$ and $A_0^{(\lamp)} \approx 2 M_{1/2}$
\end{center}
\eit {\bf BP1}, {\bf BP2} and {\bf BP3} each lie in distinct sections of the BR: stau
LSP region, neutralino LSP region dominated by three-body decays and
neutralino LSP region dominated by two--body decays,
respectively. This is depicted in Fig~\ref{TriBi}, where the ratio
between three-- and two--body decay modes of the neutralino LSP is
displayed.  The two--body $\tilde \chi^0_1$ decay modes dominate at large $M_{1/2}$ and $M_0$.
 As one can also see in this figure, the stau LSP region
within our BR is approximately given by
\beq M_{1/2}\ge 3\; M_0 - 80\,\rm{GeV},
\label{stau_lsp_region}  
\eeq 
 since the lightest neutralino mass is driven to larger values by the
large $M_{1/2}$. In general, the lighter stau mass eigenstate is
mostly right--handed. 
In Sect.~\ref{B3neutrinoMasses}, we discussed that the absolute
magnitude of the L--violating parameters as well as the relative
magnitude between them does not vary significantly with $M_0$ and
$M_{1/2}$. This implies that the LSP decay branching ratios are hardly
affected by variations of the L--violating parameters within our
BR. However, the decay modes are importantly affected by two points, as 
illustrated in Fig.~\ref{TriBi}:
\bit
\item[(A)] Whether we are in the stau or neutralino LSP region
\item[(B)] The ratio between three-- and two--body decay modes within
  the neutralino LSP
  region. 
  \eit 
  
  In the stau LSP region, the 1 and 2 lepton channels are dominant for
  large regions of parameter space. The 0 lepton channel only becomes
  significant once the stau becomes heavier than the top--quark. Then,
  hadronic stau decays via $\lamp_{i33}$ contribute significantly and
  the 1 and 2 lepton studies perform much worse, resulting in a
  ``cutoff" of the sensitive region for stau masses above the top
  mass. Now, the 0 lepton channel could further exclude parameter
  space; however, since this region extends well above $M_{1/2} \approx 500$
  GeV, we expect that the amount of data collected is not yet large
  enough to make exclusion possible.  In the neutralino LSP region
  dominated by three--body decays, we expect the 0 lepton channel to
  be the best, whereas in the case of two--body decays, the 2 lepton
  channel should perform better.

We now come to a discussion of  possible additions to the  final state particles from ``$X$"
[as contained in Eqs.~(\ref{prod_stau_lsp}) and (\ref{prod_chi_lsp})]
and the most important distributions for our benchmark region.

\begin{center}
\begin{figure}
\includegraphics[scale=0.33,angle=-90]{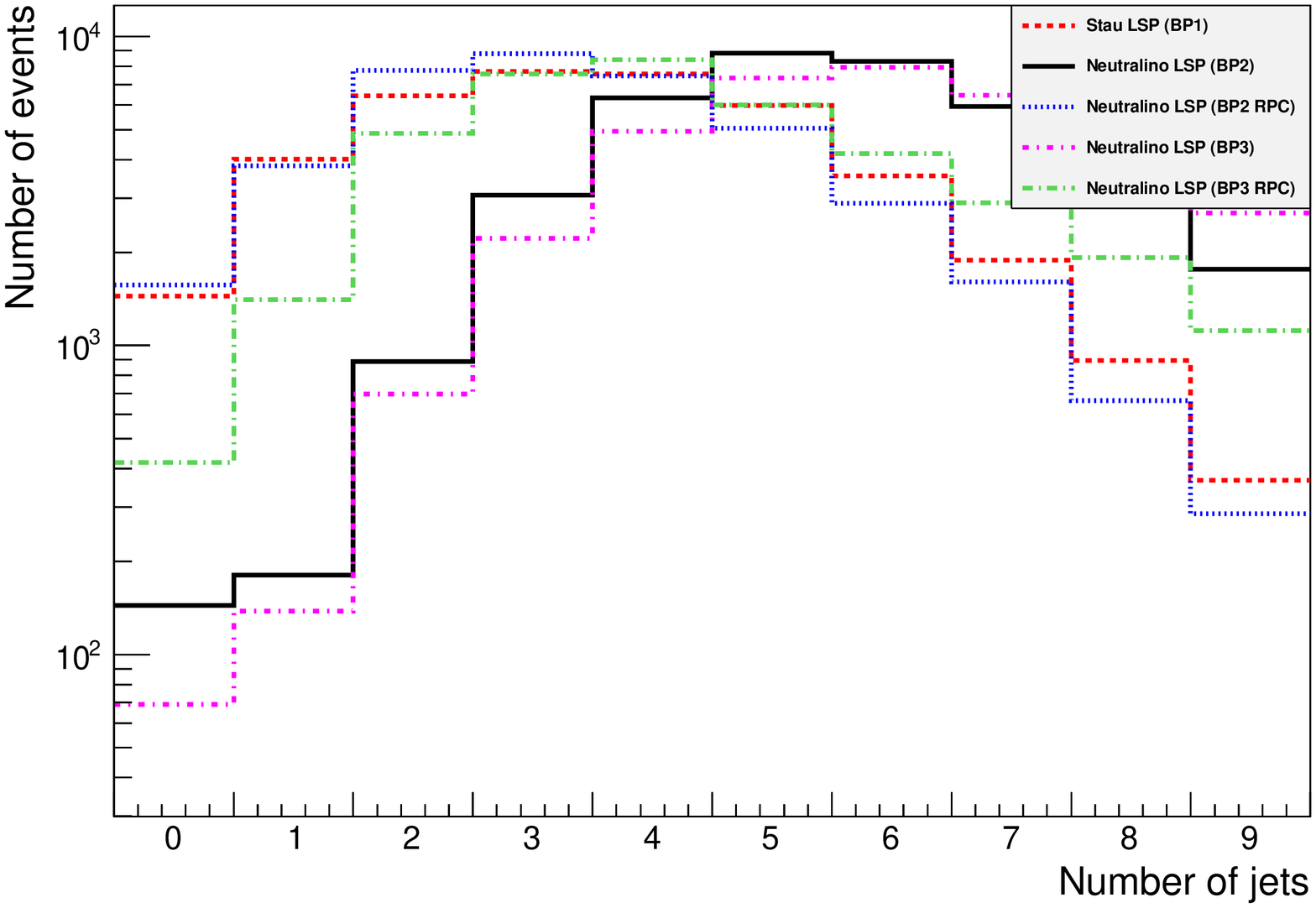} \\ \vspace{0.6cm}
\includegraphics[scale=0.33,angle=-90]{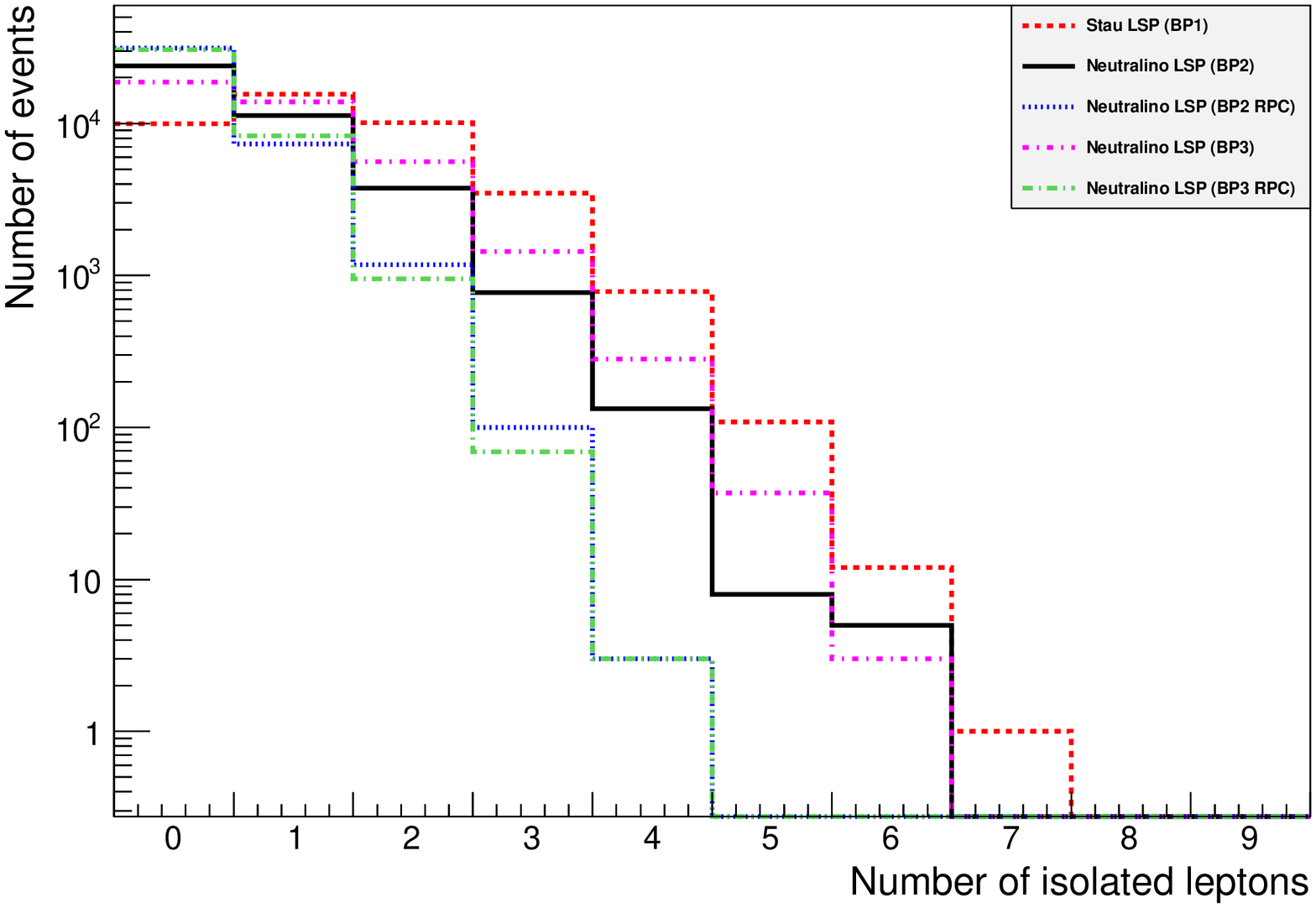} \\ \vspace{0.6cm}
\includegraphics[scale=0.34,angle=-90]{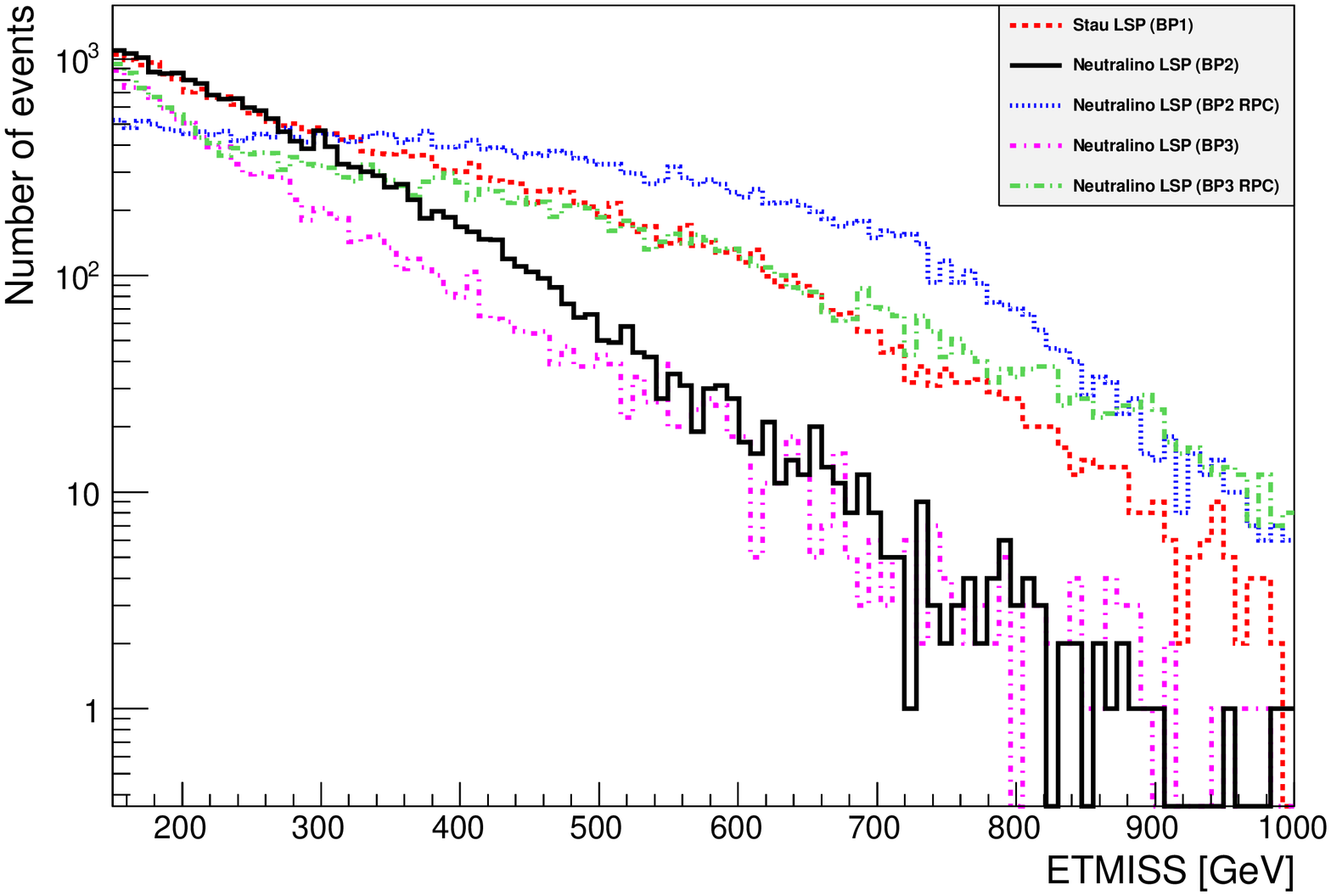}
\caption{We depict (i) the number of jets $N_{\rm jet}$, (ii) the number of isolated leptons $N_\ell$ with
  $p_T>20$ GeV and (iii) the missing transverse momentum (``ETMISS") for our
  benchmark points {\bf BP1}, {\bf BP2} and {\bf BP3}. Additionally we display an $\rp$
  version of {\bf BP2} and {\bf BP3} (``{\bf BP2 RPC}",``{\bf BP3 RPC}"), where the neutralino
  LSP is kept stable. We generated 40000
    events for each benchmark point.}
\label{stacks}
\end{figure}
\end{center}

  Additional jets can arise from gluinos in the hard process, since the
  gluino decays into quark and (virtual) squark, leading to more jets in the final state~\cite{foot5}.
Besides gluino pair and gluino--squark production, gluinos can occur in squark decays if
$M_{1/2} \ll M_0$.
   For example, in {\bf BP3}  the gluinos are lighter
  than the squarks and a sizable fraction of the squarks decay
  into a gluino and a quark which then decays via virtual squark and quark. Thus, we
    expect a higher jet multiplicity than for {\bf BP1} or  {\bf BP2}, where $m_{\tilde q} < m_{\tilde g}$. 
This is illustrated in Fig.~\ref{stacks} (i). There, we show the distribution of the
number of jets for our three benchmark points as well as for a $\rp$--conserving version of 
{\bf BP2} and {\bf BP3}  with a stable LSP (denoted ``{\bf BP2 RPC}" and ``{\bf BP3 RPC}", respectively). 
One can see that for  {\bf BP2 RPC}, there are on average only 2-3 jets
 because here squarks typically decay into a neutralino/chargino and a quark, whereas for  {\bf BP3 RPC}, there are 3-4 jets.
Comparing {\bf BP2 RPC} to  {\bf BP2}, we expect up to 4 additional b--jets from the neutralino LSP decays [Eq.~(\ref{decaysBP2})], 
and thus the distribution peaks around $N_{\rm jet} = 5-6$, cf. Fig.~\ref{stacks} (i).
Similar observations can be made for  {\bf BP3}. Here, there are more jets from the (R--parity conserving) 
 decay chain involving gluinos. However, on average there are less jets from neutralino LSP decays, Eq.~(\ref{decaysBP3}), such that the distribution also peaks at $N_{\rm jet} = 5-6$.
In the stau LSP case ({\bf BP1}), the distribution peaks at $N_{\rm jet} = 3-4$. Here there are only few jets which can be attributed to $X$ (ie. gluino decays), as discussed above. 

  Further leptons in the final state can emerge in the cascade decays of the
    SU(2) doublet squarks. The latter decay into charginos and neutralinos with dominant
    SU(2) gaugino composition, which are typically $\tilde\chi_1^\pm$
    and $\tilde\chi_2^0$ in the cMSSM.  $\tilde\chi_1^\pm$
    and $\tilde\chi_2^0$ subsequently
    decay either into slepton and lepton or gauge boson/Higgs and
    the lightest neutralino. However, this leads to isolated leptons in only $\sim\! 15\%$ of
    events in our case, as is illustrated in Fig.~\ref{stacks} (ii) by the $N_{\ell}$ distributions 
    for  {\bf BP2 RPC} and  {\bf BP3 RPC}.
   The reason for this is that in {\bf BP2}, the $\tilde\tau_1$ is
    much lighter than the other sleptons, whereas the latter are
    heavier than $\tilde\chi_2^0$ and $\tilde\chi_1^\pm$. Thus
    $\tilde\chi_2^0$ and $\tilde\chi_1^\pm$ dominantly decay into
    $\tilde\tau\tau$ and $\tilde\tau\nu$, respectively. About one third of these $\tau$'s 
    decay leptonically, leading to final state leptons. In
    {\bf BP3}, all sleptons are heavier than $\tilde\chi_1^\pm$ and
    $\tilde\chi_2^0$ and hence the latter preferably decay into a
    gauge/Higgs boson and the lightest neutralino. 
Comparing {\bf BP2 RPC} and {\bf BP3 RPC} with the corresponding $\rpv$ scenarios, we
    clearly see that there are significantly more leptons for {\bf BP2} and {\bf BP3}
      due to leptonic decays of $\tilde\chi_1^0$.
      However, there are
    more entries in the 0 lepton bin for {\bf BP2} and  {\bf BP3} than expected 
    from Eqs.~(\ref{decaysBP2}) and~(\ref{decaysBP3}),
    because some of the leptons are non-isolated or too soft or do not fall 
    into the acceptance region of the tracking system.
 The same holds for {\bf BP1}, which has overall the largest number of isolated leptons; 
 nevertheless the ratio between events with 1 lepton
 and 0 leptons is still less than predicted from Eq.~(\ref{stau_final_signature}). 
 
    In Fig.~\ref{stacks} (iii), we present the missing transverse momentum
    distribution. Here, we clearly see that {\bf BP1} has the
    hardest distribution among all $\rp$ violating distributions. Note
    that for the two other $\rp$ violating scenarios the missing
    transverse energy distribution is much softer compared to the
    respective $\rp$ conserving scenarios, due to the LSP decays.

\section{Numerical Results: Exclusion limits on hierarchical $\bt$
  cMSSM parameter space}
\label{sec:lhc_analysis} 
In this section, we further constrain the hierarchical $\bt$ cMSSM
parameter space using data from the LHC at $\sqrt{s}=7$ TeV with an
integrated luminosity of up to $5$ fb$^{-1}$.  We focus on recent
ATLAS studies with 0,1 or 2 isolated leptons, several jets and large
missing transverse momentum.  A short overview over the ATLAS studies
used is given in Table~\ref{ATLAStable}. Full details of objects
reconstruction, definitions of all kinematical observables and event
selection cuts of all three analyses can be found in the respective
ATLAS publications~ \cite{Aad:2011ib,Aad:2011qa,ConfNote0lept} ($0$
lepton),~\cite{ATLAS:2011ad,ConfNote1lept} ($1$ lepton)
and~\cite{Aad:2011cwa} ($2$ leptons). We have chosen these analyses
because they only rely on simple objects such as electrons, muons,
jets and missing transverse momentum in the final state. Thus, we do
not rely on complicated tau reconstruction and b--tagging algorithms,
which are difficult to simulate with the detector simulation {\tt
  Delphes1.9} \cite{Ovyn:2009tx}. In particular, difficulties arise in
reconstructing hadronically decaying taus~\cite{Desch:2010gi}. Also,
the published ATLAS studies for supersymmetry involving taus
\cite{tau-atlas-search} or b--jets \cite{b-atlas-search} in the final
states have smaller cross--sections or smaller efficiencies than the
multi--jet, large $\met$ and lepton searches. Thus, we expect the
``simple'' 0-2 lepton analyses to perform better with the current
amount of data.  So far, the experimental data is in agreement with
the SM background expectations. We use their results in order to
derive the $68\%$ and $95\%$ CL exclusion regions in the
$M_0$--$M_{1/2}$ parameter space.  We plan to investigate exclusion
limits arising from third generation studies and multi--lepton studies
in a future publication.

ATLAS and CMS have recently published conference notes which found
that the lightest Higgs is at least heavier than $117.5$ GeV at $95\%$
CL~\cite{atlasHiggs,Chatrchyan:2012tx}. In the hierarchical $\bt$
cMSSM, the lightest Higgs is typically rather lighter than 116 GeV,
because the value of $A_0$ is necessarily fixed to be positive and
similar in magnitude to $2 M_{1/2}$, cf.
Sect.~\ref{B3neutrinoMasses}. This means that the stop mixing cannot
become very large and thus the loop contributions to the lightest
Higgs mass are moderate. We have checked various values of $\tan\beta$
and both $\sgnmu=\pm1$; however, we found that the Higgs mass does not
become larger than 117 GeV for $M_0$, $M_{1/2} < 1$ TeV. Therefore,
the exclusion limits derived from this lightest Higgs mass bound would
by far exceed the exclusion limits derived from the 0, 1 and 2 lepton
channels mentioned above. However, it could be possible to soften the
bound if we extend the field content of the hierarchical $\bt$ MSSM by
a singlet, i.e. working in the next-to minimal SSM (NMSSM)
\cite{Ellwanger:2009dp,Vasquez:2012hn,Ellwanger:2012ke}. We leave this
topic for a future investigation at a time when there is more
certainty regarding the lightest Higgs mass.

Before applying the model independent cross section limits from the
ATLAS searches to our neutrino model, we checked that the Monte Carlo
tools are correctly tuned.  Therefore, we generated 20000 events for
each grid point in the $M_0$--$M_{1/2}$ plane in the R--parity
conserving cMSSM.  We determined the $95\%$ CL exclusion
region in the $M_0$--$M_{1/2}$ plane for the ATLAS ``1lepton-3j''
study (cf. Table~\ref{ATLAStable}) and verified that our results are
compatible with the interpretation from ATLAS within $\pm 30$ GeV.  We
now discuss the $0$, $1$ and $2$ lepton channels in detail.

\begin{table}[t]
\begin{center}
\begin{tabular}{c|ccc}
  & 0lept--SRE--m & 1lept--3j & 2lept--OS--4j \\  \hline
  $N_{\ell}$ & 0 & 1 & 2 \\
  $N_{\rm jet}$ & 6 & 3 &$ \geq 4$\\
  $p^T_{\rm jets}$ & \footnotesize $>\!\!(130,60,60,60,40,40)$ &  \footnotesize $>\!\!(100,25,25)$ & \footnotesize $>\!\!(100,70,70,70)$\\ \normalsize
  $\met$ & $>160$ & $>250$ & $>100$\\
  $ m^{\rm inc}_{\textrm{eff}}$ & $>1200$ & $>1200$ & --  \\
  $\frac{\met}{m_{\textrm{eff}}}$ & $>0.15$ & $>0.3$ & -- \\
  $\mathcal L$ & 4.7 fb$^{-1}$ & 4.7 fb$^{-1}$ & 1.0 fb$^{-1}$
\end{tabular}
\end{center}
\caption{The main cuts used in the ATLAS studies used in this collider study. 
  More details concerning the cuts can be found in the relevant ATLAS studies ($0$ lepton
  \cite{ConfNote0lept}, $1$ lepton \cite{ConfNote1lept} and $2$
  lepton \cite{Aad:2011cwa}). $N_{\ell}$ denotes the number of
  isolated leptons, $N_{\rm jet}$ the number of
  jets and $p^T_{\rm jets}$ specifies the minimal transverse momentum 
  which is required for these jets.  $\met$ gives the minimal
  value of missing transverse momentum
  of the event, $ m^{(inc)}_{\textrm{eff}}$ the minimal (inclusive) effective mass
  and  $\mathcal L$ denotes the total integrated luminosity at 7 TeV. }
\label{ATLAStable}
\end{table}%

\subsection{0 lepton channel}
ATLAS has used the 0 lepton channel as one of the first search
channels for supersymmetry
\cite{Aad:2011ib,Aad:2011qa,ConfNote0lept}. So far, they have
collected a total luminosity of about 4.7 fb$^{-1}$ at the center of
mass energy of $\sqrt{s}=7$ TeV. From the non--observation of an
excess, we can derive exclusion limits on the hierarchical $\bt$
cMSSM.  The ATLAS 0 lepton channel is divided into several signal
regions (SR). For all signal regions, the cut on $\met$ and the
minimum requirement on $p^T_{\rm jet}$ of the first two
most--energetic jets are identical. However, the number of jets and
the minimum $p^T_{\rm jet}$ cut for the remaining jets as well as the cut
on $m^{inc}_{\textrm{eff}}$ and on the ratio $\met / m_{\textrm{eff}}$
differ for the different signal regions.

\begin{center}
\begin{figure}
\includegraphics[scale=0.6]{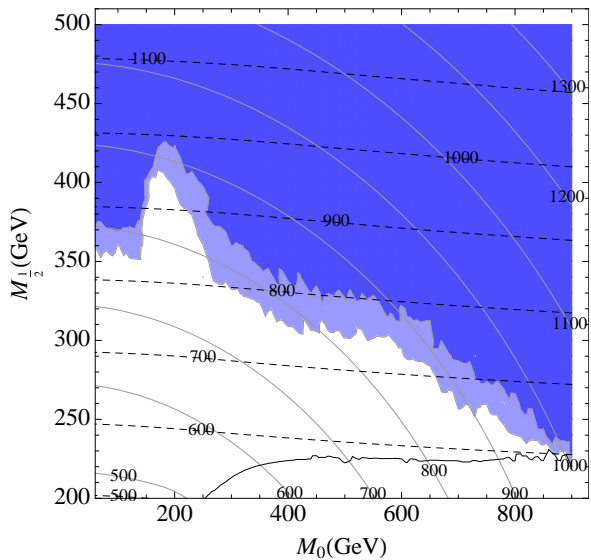}
\caption{Exclusion limit on our benchmark region, where
  $\tan\beta=25$, $\sgnmu=1$ and $A_0^{(\lamp)} \approx 2 M_{1/2}$, from the 0 isolated
  leptons, 6--jets and MET (``0lept--SRE--m") ATLAS study. The white
  region is excluded at $95\%$ confidence level (CL), the light blue
  is excluded at $68\%$ CL. The grey lines denote the squark masses, 
  the dashed black
  lines denote the gluino masses (each in GeV).  
The black line delineates the region (below) where the
  lifetime of the LSP becomes larger than $c \tau \gtrsim 15$ mm. 
 In this region, the exclusion limits are not applicable because the 
 ATLAS study rejects leptons and jets which do not originate from the
 primary vertex.}
\label{exclusion_0lep}
\end{figure}
\end{center}

We have examined all signal regions after applying the object
reconstruction described in their study and found that we obtain the
strictest exclusion limits for the ``SRE--m" signal region, which
demands 6 jets, $m^{\rm incl}_{\textrm{eff}} >1200$ GeV and
$\frac{\met}{m_{\textrm{eff}}}>0.15$, cf. Table~\ref{ATLAStable}.  We
show the resulting plot in the $M_0$--$M_{1/2}$ plane in
Fig.~\ref{exclusion_0lep}.  The exclusion limit peaks at $M_0 \approx
200$ GeV.  This is the region where the neutralino LSP decays
dominantly via three--body decays $\tilde\chi_1^0\rightarrow \nu b\bar
b$, {\it c.~f.}  Fig.~\ref{TriBi}. It was to be expected that the
``SRE--m" signal region gives good exclusion limits for this type of
scenario, because if both neutralinos decay via
$\tilde\chi_1^0\rightarrow \nu b\bar b$, we expect at least 6 parton
level jets (including b--jets).  Also, we have only moderate $\met$
because of the three--body decay of the neutralino, and therefore more
events survive in the ``SRE--m" than in the ``SRE--t" scenario (where
$m^{inc}_{\textrm{eff}} >1500$ GeV).  Finally, leptons from the
cascade decays of SU(2) doublet squarks into $\tilde \chi_1^\pm$ and
$\tilde \chi_0^2$ are suppressed, since the latter dominantly decay
into $\tilde \chi_1^\pm\rightarrow \tilde\tau \nu$ and
$\tilde\chi_2^0\rightarrow \tilde \tau\tau$.

For increasing $M_0$, the
exclusion region decreases to lower $M_{1/2}$ values. We can see in
Fig.~\ref{TriBi} that the two--body decay mode of the neutralino
becomes more important here. Thus, an increasing number of the
neutralino LSPs decay into a gauge boson and a lepton and less
$b$--jets are expected in the final state, so that less events pass
the kinematical cuts on the final state jets. Another effect is that
for larger $M_{0}$, the production cross section decreases. 

Directly to the left of the peak at $M_0 \approx 200$ GeV, the limit
drops off sharply because here the LSP becomes the $\tilde \tau_1$ and
there are significantly less events with 6 jets and no
leptons. However, $M_{1/2} \lesssim 350$ GeV can still be excluded at
$95 \%$ CL. We would like to point out that in principle, it is
possible to obtain better exclusion limits (up to $M_{1/2} \lesssim
400$ GeV) in the stau LSP case by using a signal region with only 4 or
5 jets. However, the 1 lepton study performs even better and therefore
we go not into detail about the results from these signal regions
here.

We do not consider the region with a LSP lifetime exceeding
  $c\tau=15$ mm, since the ATLAS searches for supersymmetry require
  prompt LSP decays. In Fig. \ref{exclusion_0lep}, the region below
  the solid black curve highlights a LSP with a lifetime $c\tau\ge15$
  mm.

\subsection{1 lepton channel}
Refs.~\cite{ATLAS:2011ad, ConfNote1lept} search for multi--jet events
with large missing transverse momentum and exactly one isolated
lepton. Similarly to the 0 lepton channel in the previous subsection,
the 1 lepton channel was one of the first supersymmetry search
channels and the current integrated luminosity is $4.7$ fb$^{-1}$ at
the center of mass energy of $7$ TeV. They consider signal regions
with 3-- or 4--jets with different kinematic configurations, which are
optimized for the $\rp$ cMSSM with a large mass difference between the
gluino and the LSP.  Additionally, they include a soft--lepton signal
region which is sensitive to scenarios with small mass splitting
between the sparticles.

\begin{center}
\begin{figure}
\includegraphics[scale=0.6]{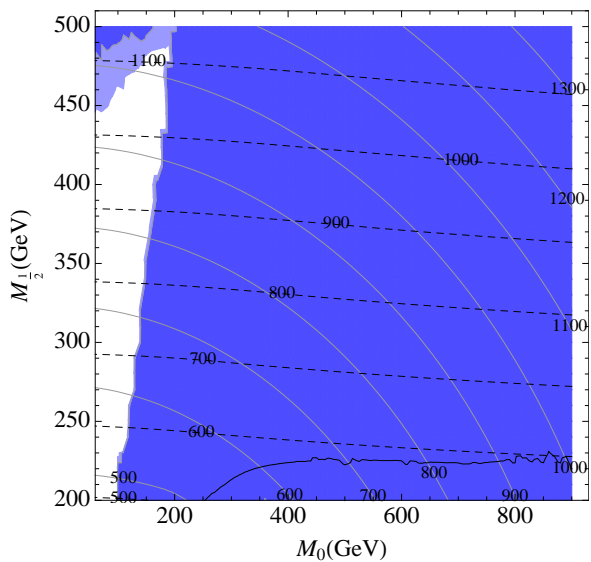}
\caption{Exclusion limit on our benchmark region, where
  $\tan\beta=25$, $\sgnmu=1$ and $A_0^{(\lamp)} \approx 2 M_{1/2}$, 
  from the 1 isolated
  lepton, 3--jets and MET (``1lept--3j") ATLAS
  study~\cite{ConfNote1lept}. The white region is excluded at $95\%$
 CL, the light blue is excluded at $68\%$ CL. 
  The grey lines denote the squark masses, 
  the dashed black
  lines denote the gluino masses (each in GeV). 
The black line delineates the region (below) where the
  lifetime of the LSP becomes larger than $c \tau \gtrsim 15$ mm. 
   In this region, the exclusion limits are not applicable because the 
 ATLAS study rejects leptons and jets which do not originate from the
 primary vertex. }
\label{exclusion_1lep}
\end{figure}
\end{center}

Comparing the results for the different signal regions, we observe
that the 3--jet signal region (``1lept--3j'' ) provides us with the
best overall exclusion limits in the stau LSP region up to
$M_{1/2}\sim500$ GeV (i.e. better than the limits from any other
signal region in the 0 to 2 lepton channels).  The main kinematic cuts
of the 1lept--3j signal region are listed in Table~\ref{ATLAStable}
and the resulting plot is shown in Fig.~\ref{exclusion_1lep}. Almost
half of the events in the stau LSP region decay into final states with
1 lepton, cf. Sect.~\ref{stauSignatures}.  Note also that the 1 lepton
study~\cite{ConfNote1lept} demands the most stringent cut on $\met$
among the 0, 1 and 2 lepton studies.  In the stau LSP region with
direct (two--body) leptonic decays, much more missing transverse
momentum is produced than in the neutralino LSP region. In particular
in the neutralino LSP region with dominant three--body decays into
$\nu \bar b b$, the amount of $\met$ is greatly reduced compared to
the stau LSP region. Moreover, much less charged leptons arise from
the neutralino decay. Additional leptons from the cascade decays are
also heavily suppressed. Thus, we have a sharp drop of the acceptance
in the crossover region between the stau and neutralino LSP region.
For larger $M_0$ values, eventually the two--body neutralino decay
modes become dominant over the three--body decay mode. However, the
hard cut on $\met$ still rejects many signal events in this region.

Note that the ATLAS signal region with 1 lepton and 4--jets is also
sensitive to the neutralino LSP region besides the stau LSP region.
This explains why in the old 4--jet signal region with 1 fb$^{-1}$ in
the muon channel, ATLAS was able to constrain the bilinear $\rpv$
model presented in Ref.~\cite{ATLAS:2011ad} (with two--body neutralino
decays) quite well. However, having in mind that in our case we have
addtional three--body decays and in the new 5 fb$^{-1}$ study, the
cuts are more stringent cuts than the 1 fb$^{-1}$ version and not
optimized for our type of scenario, the resulting exclusion limits on
the neutralino LSP region are weaker than the limits derived in the 2
lepton channel as shown below.

\subsection{2 lepton channel}
The ATLAS study based on final states with two leptons and missing
transverse momentum \cite{Aad:2011cwa} has not yet been updated to
include more than 1 fb$^{-1}$ of data. The search is divided into
opposite--sign (OS), same--sign (SS) and flavour--subtraction (FS)
signal regions where up to 4 jets are demanded besides exactly 2
leptons and a cut on $\met$.  We find that we obtain the best
exclusion limits with the OS signal regions. The three OS regions
differ in the $\met$ cut, the number of jets and the corresponding
minimal $p^T_{\rm jets}$ cut.  As in the case of the 1 lepton channel,
the OS studies with the hardest transverse missing momentum cut
(``2lept--OS--2j", $\met>250$ GeV) are quite sensitive to the stau LSP
region where two staus decay leptonically. However, in the 2 lepton
channel the obtained exclusion limits are $\sim\!50$ GeV weaker than
in the ``1lept--3j'' study.  This is due to the stringent cuts on
$m^{inc}_{\textrm{eff}}$ and on the ratio $\met / m_{\textrm{eff}}$ in
the ``1lept--3j'' search channel, which yield better signal isolation
and background suppression.

\begin{center}
\begin{figure}
\includegraphics[scale=0.6]{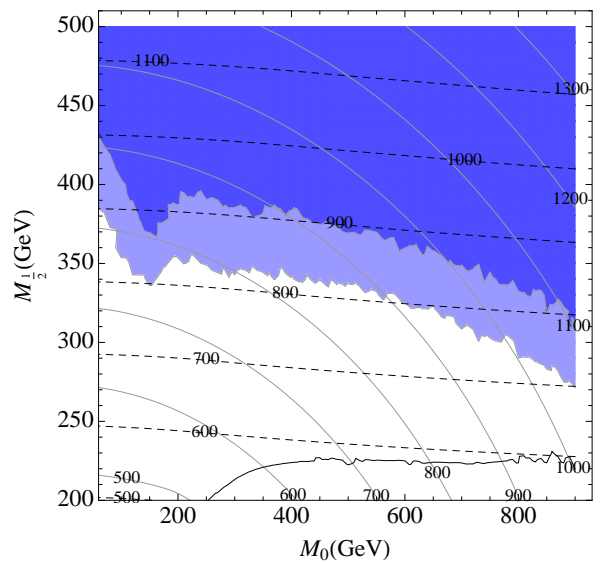}
\caption{Exclusion limit on our benchmark region, where
  $\tan\beta=25$, $\sgnmu=1$ and $A_0^{(\lamp)} \approx 2 M_{1/2}$, from the 2 isolated
  opposite--sign leptons, 4--jets and MET (``2lept--OS--4j") ATLAS
  study. The white region is excluded at $95\%$ CL,
  the light blue is excluded at $68\%$ CL. The grey lines denote the squark masses, 
  the dashed black
  lines denote the gluino masses (each in GeV). 
  The black line delineates the region (below) where the
  lifetime of the LSP becomes larger than $c \tau \gtrsim 15$ mm.
   In this region, the exclusion limits are not applicable because the 
 ATLAS study rejects leptons and jets which do not originate from the
 primary vertex.}
\label{ratioTriBi}
\label{exclusion_2lep}
\end{figure}
\end{center}

 The OS and 4--jet channel with a moderate $\met$ cut of 100 GEV
 (``2lept--OS--4j''), described in Table~\ref{ATLAStable}, provides us
 with the best exclusion limits for $M_0 \gtrsim 300$ GeV, where the
 neutralino LSP decays dominantly via two--body decays as shown in
 Fig.~\ref{exclusion_2lep}.  We notice a slight dip for smaller $M_0$
 ($M_0 \!\sim 200$ GeV), where there are dominant three--body
 neutralino decays. Here, as discussed in the previous subsections,
 parton--level leptons from the neutralino LSP decays or from the
 cascade decays of the SU(2) doublet squarks are heavily suppressed
 and the exclusion limits from the 0 lepton channel are more
 stringent.  For even smaller values of $M_0$, we are in the stau LSP
 region and the exclusion limits improve again.  However, as discussed
 in the last paragraph, the cuts are not optimized for a stau LSP
 scenario.  The $E_T^{\rm miss}$ cut is the weakest among all three
 analyses in Table~\ref{ATLAStable} and the kinematic requirements on
 the jets are harder compared to the ``1lept--3j'' search channel.

For $M_{0}\gg M_{1/2}$, the gluino is
generally lighter than the squarks and thus we expect a higher jet
multiplicity and in general more jets passing the kinematic cuts.
However, much less transverse momentum is generated compared to the
R--parity conserving case or the stau LSP region. Thus, the
``2lept--OS--4j'' yields the better overall exclusion region in the
neutralino LSP region with dominant bilinear RPV decays due to the
softer $E_T^{\rm miss}$ cut compared to "0lept--SREm". One further remark
on the number of leptons in the final state: for $M_{1/2}\ll M_0$, the
SU(2) doublet squarks decay via a wino--like gaugino is quite sizable,
although we have the competing decay channel via an off--shell
gluino. These wino--like gauginos again dominantly decay into gauge
bosons providing additional leptons in the final state.
\section{Summary and Conclusion}\label{sec:summary} 
We introduced a hierarchical ansatz for the L--violating trilinear
Yukawa couplings in the $\bt$ cMSSM. Here, the trilinear LNV Yukawa
couplings are related to the Higgs Yukawa couplings via six
independent parameters $\ell_i$ and $\ell_i^\prime$. We have then
determined the best fit values of the $\ell_i$ and $\ell_i^\prime$ in
order to obtain phenomenologically viable neutrino masses and mixing
angles.  It is possible to quasi unambiguously determine the
L--violating sector as well as the value of the SUSY breaking scalar
coupling $A_0$ from neutrino oscillation data.  We discussed the final
collider signatures in the stau LSP and neutralino LSP scenarios at
the LHC and finally used the ATLAS searches in jets and large missing
transverse momentum with 0, 1 and 2 isolated leptons in order to find
the $95\%$ and $68\%$ CL exclusion limits in the $M_0$--$M_{1/2}$
plane for fixed $\sgnmu$ and $\tan\beta$. We can exclude squark masses
below 800 GeV, and gluino masses below 700 GeV (for squark masses
below 1 TeV) at $95\%$. These limits become more stringent at $68\%$
CL, by roughly 100 GeV. Compared to the case of the R--parity
conserving cMSSM, we obtain weaker limits because generally we have
more jets and leptons and less $\met$ due to the LSP decays.
  
We want to conclude with a short discussion of how we can improve a
future collider study for our model or similiar R--parity violating
models.  There are a number of studies in which R--parity violating
collider signatures are investigated, as mentioned in the
introduction.  Many of these studies consider multilepton
($N_\ell\ge3$) signatures in association with much less missing
transverse energy than in our study. They typically assume, however, a
single non--zero $\lambda_{ijk}$ without third generation indices,
ie. $i,j,k\in\{1,2\}$, so that the number of lepton is enhanced. In
our model, the LSP dominantly decays via $\lam_{i33}$ or
$\lambda_{i33}^\prime$ couplings involving third generation decay
products, or via neutralino--neutrino mixing involving gauge boson
decay products.  However, the branching ratio of the LSP into leptons
is still considerably large (between $19\%$ and $47\%$) and therefore
the lepton multiplicity is higher than in R--parity conserving
models. Also, the average $p^T_{\ell}$ distribution of the signal
leptons will be relatively hard due to the large phase space of the
two body decay channels of the LSP into SM fermions. For example, in
{\bf BP2} the hardest lepton has on average $p_{\ell}^{T}=80$ GeV. In
{\bf BP2 RPC}, the hardest lepton has a mean value of
$p_{\ell}^{T}=60$ GeV. Demanding one (two) lepton(s) with moderate
$p^T_{\ell}$ cuts might be advantageous to isolate the signal. As an
alternative, we can also apply a kinematical cut on the scalar sum of
all the leptons' $p^T_{\ell}$.

Decays via trilinear couplings with third generation indices are
dominant in large regions of parameter space in our model. Therefore,
we expect a substantial proportion of events with third generation SM
particles in this parameter region. For example, we expect a large
number of taus and b--jets in {\bf BC1} and {\bf BC2},
respectively. In {\bf BP2}, about $50\%$ of all events have at least
one b--jet. This is in sharp contrast to {\bf BP2 RPC} where only
$13\%$ of all events have a b--jet in the final state.  Requiring
hadronically decaying taus or b--jets in the final state should help
to suppress the SM background. However, for the parameter region
around {\bf BP3}, the LSP dominantly decays via neutralino--neutrino
mixing. Here, we do not expect third generation particles in the final
state in abundance.

The increase in jet and lepton multiplicities due to LSP decays in our
model happens at the cost of less missing transverse momentum compared
to the R--parity conserving case. For example, in {\bf BP3 RPC} we
have on average $\mpt=213$ GeV because the stable neutralino LSP
escapes detection. In {\bf BP3} we obtain a mean value of $\mpt=123$
GeV due to neutrinos from the LSP decay. 
%
In many studies the effective mass,
\begin{equation}
M_{\rm eff}=\mpt+\sum p^T_{\rm jets},
\label{eq:mass_eff}
\end{equation}
is used to ``measure'' the effective SUSY mass scale. However, they
assume a stable LSP and thus $M_{\rm eff}$ receives a sizable
contribution from $\mpt$.  Our
signatures tend to look softer than those of most R--parity conserving scenarios 
%
%
because some of the decay products of the LSP 
are not included in the sum in Eq.~(\ref{eq:mass_eff}). 
A useful discriminating variable to increase the
significance of our signal could be the scalar sum of missing transverse
momentum, all jets, leptons and hadronic taus,
\begin{equation}
S_T=\mpt+\sum p^T_{\rm jets}+\sum p^T_{\ell}+\sum p_{\tau_{\rm
    had}}^T\,.
\label{eq:mass_eff_rpv}
\end{equation}
 For example, the ratio
of Eq.~(\ref{eq:mass_eff}) and Eq.~(\ref{eq:mass_eff_rpv}) is 0.85 for
{\bf BP3}.

Finally, it is difficult to constrain the region $M_{1/2}\lesssim 230$
GeV in our model due to the finite lifetime of the LSP, since many
supersymmetry searches only reconstruct leptons and jets which
originate from the primary vertex.  We thus conclude that allowing
events with displaced vertices would certainly be advantageous to
establish bounds in the low $M_{1/2}$ region.

 
\begin{acknowledgments} 
  We thank H.K. Dreiner, C.-H. Kom and  A. Williams for useful discussions. 
  J.S.K. thanks the University
  of Bonn and the Bethe Center for Theoretical Physics for hospitality
  during numerous visits. M.H. thanks the University of Adelaide for
  hospitality during her visit.   This work is supported in part by the
  Deutsche Telekom Stiftung, by the Bonn-Cologne Graduate School of
  Physics and by the ARC Centre of Excellence for Particle Physics at the
  Terascale.
 
\end{acknowledgments} 
 
\bibliographystyle{h-physrev}

\begin{thebibliography}{99}
\bibitem{foot1}
  Because of the
  antisymmetry of $\lam_{ijk}$, $\lam_{333}=0$ and $\ell_3$ could only
  contribute to neutrino masses via $\lambda_{233}$.  This means that
  for a sizable contribution, $\ell_3$ must be several orders of
  magnitude larger than $\ell_1$ or $\ell_2$.
\bibitem{foot2}
  $\ell_3$ has no relevance for the
  collider signatures as long as it doesn't become several orders of
  magnitude larger than $\ell_1$ and $\ell_2$.
\bibitem{foot4} 
  Only in a small part of the neutralino
LSP region, where $M_{1/2} \lesssim 240$ GeV, the lifetime of the LSP can become larger than
 $c \tau \gtrsim 15$ mm.
\bibitem{foot5}
Note that additional jets can also arise from QCD Bremsstrahlung.
\end{thebibliography}

 
\end{document}